\documentstyle[epsf]{mn}

\def\RCS$#1:#2${\expandafter\def\csname RCS#1\endcsname{#2}}
\RCS$Revision: 1.16 $ \RCS$Date: 2000/01/20 12:26:54 $

\title[Angle-averaged microlensing amplification
	functions]{Gravitational microlensing source limb-darkening
	and limb-polarization,~I: angle-averaged amplification functions}
\author[Norman Gray]{Norman Gray\newauthor Department of Physics and
	Astronomy, University of Glasgow, Glasgow G12 8QQ, UK}
\date{\RCSDate}

\newcommand{\be}[1]{\begin{equation}\label{e:#1}}
\newcommand{\ee}{\end{equation}}
\newcommand{\bea}{\begin{eqnarray}}
\newcommand{\eea}{\end{eqnarray}}
\newcommand{\bean}{\begin{eqnarray*}}
\newcommand{\eean}{\end{eqnarray*}}
\newcommand{\xb}{\overline x}
\newcommand{\I}[1]{\tilde I^{(#1)}}
\newcommand{\Q}[1]{\tilde Q^{(#1)}}

\newcommand{\dd}{{\mathrm d}}		
\newcommand{\ddd}{\,{\mathrm{d}}}
\newcommand{\e}[1]{{\mathrm{e}^{#1}}}
\newcommand{\eqnref}[1]{Eqn.~(\ref{e:#1})}
\newcommand{\figref}[1]{Fig.~\ref{f:#1}}
\newcommand{\secref}[1]{Sect.~\ref{s:#1}}

\newcommand{\RJ}{{R_J}}
\newcommand{\RD}{R_D}
\newcommand{\RF}{R_F}

\begin{document}
\maketitle

\begin{abstract}
There is increasing interest in extended source effects in
microlensing events, as probes of the unresolved sources.  Previous
work has either presumed a uniform source, or else required an
approximate or numerical treatment of the amplification function
averaged over the source disk.  In this paper, I present analytic
expressions for the angle-averaged amplification functions for the
rotationally-symmetric intensity and polarization cases.

These integrals will allow us to use the technology of inverse
problems to study the source limb-darkening and limb-polarization functions.
\end{abstract}


\section{Introduction\label{s:intro}}

The most common interest in microlensing events is as a probe of the
lensing object itself.  Recently, however, there has been increasing
interest in such events as probes of the otherwise unresolved
sources~\cite{valls-gabaud97,mao98,gaudi98}.  When the lens transits the
source (or nearly so), it breaks any rotational symmetry, and this
gives us access to the surface brightness, as well as
polarization~\cite{simmons95,simmons95a} and
chromaticity~\cite{valls-gabaud95,valls-gabaud97} information.  Previous
work on extended source effects has concentrated on the forward
problem and generally either performed the required
calculations numerically rather than analytically, or used an
approximate form of the amplification function.  Also, much of the
work on source effects has relied on the high amplification provided
by binary lens causic crossings, rather than the amplification of a
single lens.

The gravitational lensing forward problem -- that of predicting
centroid motion and magnification for a given set of source parameters
-- is relatively easy.  The problem is also, however, typically poorly
conditioned, in the sense that there will be a broad range of
limb-darkening or limb-polarization functions which could plausibly
correspond to the observed signal in a microlensing event.  This means
that a parameter-fitting approach to recovering these functions is
very dangerous.

We can make progress by expressing the problem explicitly as an
inverse problem~(IP), and using the technology of IP methods to analyse
precisely what information can be recovered for a given set of
observations.

That is the subject of, and motivation for, a forthcoming
paper~\cite{gray00a}; here I am concerned with identifying the
angle-averaged amplification functions as IP kernels, and obtaining
analytic expressions for them.  As well as facilitating the IP
analysis of the problem, these analytic kernels can help in the
treatment of the forward problem, since they can be evaluated more
efficiently than by a numerical integration, and with high accuracy
over their entire domain.

In \secref{kernels}, I define the amplification functions as IP
kernels, in \secref{int}, I integrate them about the source's centre,
and in \secref{elliptic}, I present the results of that integration.

\section{Amplification functions as inverse problem kernels}
\label{s:kernels}

\begin{figure}
\begin{center}
\epsfbox{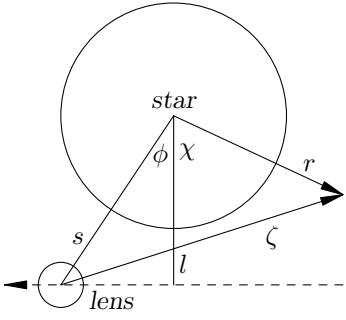}
\end{center}
\caption{\label{f:geom}Geometry of a lensing event.  The projected
path of the lens has impact parameter~$l$, and the path is
parameterised by polar coordinates~$s(t)$ and~$\phi(t)$, relative to
the centre of the source, projected into the lens plane.  Any point in
that plane can be given in polar coordinates~$r$ and~$\chi$, and this
point is a distance~$\zeta$ from the centre of the lens.  All
dimensions are normalised to the Einstein radius in the source plane.
The angles~$\phi$ and~$\chi$ are taken with respect to the line
joining the source to the lens' point of closest approach.}
\end{figure}

The geometry of a microlensing event is as shown in \figref{geom}.

The gravitational lens amplification function is~\cite{schneider92}
\be{lenseq}
	A(\xi) = \frac12\left(\xi+\frac1\xi\right),
\ee
where
\[
	\xi = \left(1+\frac4{\zeta^2}\right)^{1/2},\qquad
	\zeta^2 = r^2 + s^2 - 2rs\cos(\chi-\phi).
\]

Denote the intensity by~$I(r)$ and the Stokes parameter
by~$Q(r,\chi)=-P(r)\cos2\chi$, where~$P(r)$ is the polarization of the
stellar surface, and we are assuming that the surface brightness is
rotationally symmetric.  In the case of a microlensing event, we
cannot resolve details of the lensed source, and must therefore
measure integrals over the source surface.  We immediately obtain
\bea
I(s(t),\phi(t)) &=& \int_0^\infty I(r)\tilde A_I(r;s,\phi)\ddd r
	\label{e:IIP}\\
Q(s(t),\phi(t)) &=& \int_0^\infty P(r)\tilde A_Q(r;s,\phi)\ddd r
	\label{e:QIP}
\eea
where the amplification kernels are
\bea
\tilde A_I &=& r\int_0^{2\pi} A(r,\chi;s,\phi)\ddd \chi 
	\label{e:Idef}\\
\tilde A_Q &=& -r\int_0^{2\pi} \cos2\chi A(r,\chi;s,\phi)\ddd\chi.
	\label{e:Qdef}
\eea
Note that the kernel~$\tilde A_I$ is a factor~$2\pi r$ times the
angular average of the amplification function, and the
functions~$I(s,\phi)$ and~$Q(s,\phi)$ have the dimensions of flux
rather than intensity.

Equations~(\ref{e:IIP}) and~(\ref{e:QIP}) are in the form of an
inverse problem.  We address the inverse problem in a forthcoming
paper~\cite{gray00a}.  The evaluation of the integrals~$\tilde
A_{I,Q}$ is rather hard, and I describe it in this paper.

\section{Integration of the amplification functions}
\label{s:int}

Write
\be{z-def}
	z = \exp i(\chi-\phi),\qquad \dd z = iz\dd \chi,
\ee
so that
\[
	\zeta^2 = r^2 + s^2 - rs\left(z+\frac1z\right).
\]
Define
\be{ai-def}
	a_1 = \frac{r^2+s^2}{2rs},\qquad a_2 = \frac{4+r^2+s^2}{2rs}
\ee
and
\bea
	x_1 &=& a_1 + \sqrt{a_1^2 - 1} = 
		\left\{\begin{array}{cl}
			r/s&, r\ge s\\
			s/r&, r<   s
		\end{array}\right.
		\label{e:x1-def}\\
	x_2 &=& a_2 + \sqrt{a_2^2 - 1}
		\label{e:x2-def}.
\eea
Defining $\xb_i=1/x_i$, we have
\[
	x_i + \xb_i = 2a_i,\qquad x_i \xb_i = 1.
\]
It is easy to show that
\[
	0 < \xb_2 < \xb_1 \le 1 \le x_1 < x_2.
\]
Rewriting the expression for~$\xi$, we obtain
\be{xiz}
	\xi \equiv \left(1+\frac4{\zeta^2}\right)^{1/2}
		= \left(\frac{(z-x_2)(z-\xb_2)}{(z-x_1)(z-\xb_1)}\right)^{1/2}.
\ee

\begin{figure}
\begin{center}
\epsfxsize\hsize
\epsfbox{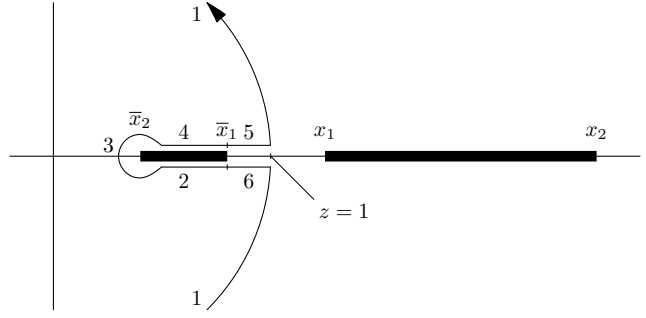}
\end{center}
\caption{\label{f:contour}The contour for the integral in
\eqnref{tI-def}, showing the cuts between the poles at~$\xb_2$,
$\xb_1$, $x_1$ and~$x_2$.}
\end{figure}

We now evaluate the integral
\be{tI-def}
	\tilde I = \oint_C \frac1z\left( \xi + \frac1\xi\right)\ddd z,
\ee
where~$C$ is the contour shown in \figref{contour}.  The integrand has
poles at~$z=x_{1,2}$, $z=\xb_{1,2}$ and~$z=0$.

The contour encloses a single singularity at~$z=0$.  We have
$\xi(z=0)=1$, so that
\be{Ival}
	\tilde I = 2\pi i \times \mathop{\mathrm{Res}}(\tilde I(z=0)) = 4\pi i.
\ee

For contour~1, substitute $z=\e{i\psi}$, for $\psi\in[0,2\pi]$.  Then
$\I1=i\int_0^{2\pi}(\xi + 1/\xi)\ddd\psi$, and the substitution
$\psi=\chi-\phi$ produces
\be{I1-res}
	\I1 = 2i\int_\phi^{\phi+2\pi} A\bigl(\xi(\chi,\phi)\bigr)\ddd\chi
		= \frac{2i}r \tilde A_I(r;s,\phi).
\ee

Contours~5 and~6 cancel, and with the substitution
$z=\xb_2+\rho\e{i\theta}$, it is clear that $\I3\to 0$ as~$\rho\to0$.

Now turn to contours~2 and~4.  By substituting $z=\xb_2+\sigma$, 
$\sigma=0\to(\xb_1-\xb_2)$, into $\I4$, substituting
$z=\xb_2+\sigma\e{i2\pi}$, $\sigma=(\xb_1-\xb_2)$, into $\I2$, and
noting that $0<\sigma<\xb_1-\xb_2$ (so that $|\sigma|<|\xb_2-x_2|$,
$|\xb_2-x_1|$, and $|\xb_2-\xb_1|$, so that the phases of the
corresponding square-rooted factors in~$\I2$ are unaffected by the
factor of $\e{i2\pi}$), we can see that
\be{I24-res}
	\I2 = \I4.
\ee
Now substituting $z=x$ ($x$ real) directly into \eqnref{tI-def}, we
obtain
\be{I4-res}
	\I4 = -i (I_1 - I_2),
\ee
where
\bea
	I_1 &=& \int_{\xb_2}^{\xb_1}\frac1x
		\left(\frac{(x_2-x)(x-\xb_2)}{(x_1-x)(\xb_1-x)}\right)^{1/2}
		\ddd x
		\label{I1-def-orig}\\
	I_2 &=& \int_{\xb_2}^{\xb_1}\frac1x
		\left(\frac{(x_2-x)(x-\xb_2)}{(x_1-x)(\xb_1-x)}\right)^{-1/2}
		\ddd x.
		\label{I2-def-orig}
\eea
Using the notation of \cite{carlson88},
\[
	[p_1,\dots,p_n] 
		\equiv \int\prod_{i=1}^n(a_i+b_it)^{p_i/2}\ddd t,
\]
we may rewrite these as
\bea
	I_1 &=& [+1,-1,-1,+1,-2]	\label{e:I1-def}\\
	I_2 &=& [-1,+1,+1,-1,-2]	\label{e:I2-def}
\eea
with $a_i=(-\xb_2,\xb_1, x_1, x_2, 0)$ and $b_i=(+1,-1,-1,-1,+1)$ in
both cases.
We will evaluate these integrals when we return to them below, in
\secref{elliptic}.

Thus, collecting Eqns.~(\ref{e:Ival}), (\ref{e:I1-res}),
(\ref{e:I24-res}) and~(\ref{e:I4-res}), we obtain the angle-averaged
amplification function as
\be{AI-res}
	\tilde A_I (r;s) = r(2\pi + I_1 - I_2).
\ee
This has been confirmed by direct numerical integration of the
integrand.

Turning now to \eqnref{Qdef}, we may again substitute
$z=\exp{i(\chi-\phi)}$, and obtain
\[
	\cos2\chi = \frac12\left(
		z^2 \e{i2\phi} + z^{-2}\e{-i2\phi} \right).
\]
Now evaluate the integral
\be{tQ-def}
	\tilde Q = \oint_C
		\left(z \e{i2\phi} + z^{-3}\e{-i2\phi} \right)
		\left(\xi + \frac1\xi\right)\ddd z,
\ee
with the same contour as above, and with~$\xi$ as in \eqnref{xiz}.
This has a third-order pole at~$z=0$,
which means that the residue is
\be{resQ}
	\mathop{\mathrm{Res}}(\tilde Q) = a_{-1}
		= \frac12\frac{\dd^2}{\dd z^2}\left[\xi+\frac1\xi\right]_{z=0}
		= (a_1 - a_2)^2
\ee
with~$a_i$ as defined in \eqnref{ai-def} above.  Thus
\be{Qval}
	\tilde Q = 2\pi i\e{-i2\phi}(a_1 - a_2)^2.
\ee

Much of the calculation goes through as before.
Substituting~$z=\e{i\psi}$, we obtain,
\be{Q1-res}
	\Q1 = -\frac{4i}r \tilde A_Q(r;s,\phi).
\ee
Contours~5 and~6 cancel, and contour~3 makes zero contribution in
the~$\rho\to0$ limit.  Similarly,
\be{Q4-res}
	\Q2 = \Q4 = i\e{ i2\phi}(-Q_1 + Q_2) 
		  + i\e{-i2\phi}(-Q_3 + Q_4),
\ee
where
\bea
	Q_1 &=& [+1,-1,-1,+1,+2]	\label{e:Q1-def}\\
	Q_2 &=& [-1,+1,+1,-1,+2]	\label{e:Q2-def}\\
	Q_3 &=& [+1,-1,-1,+1,-6]	\label{e:Q3-def}\\
	Q_4 &=& [-1,+1,+1,-1,-6]	\label{e:Q4-def}
\eea
with~$a_i$ and~$b_i$ as above.

Assembling equations~(\ref{e:Qval}), (\ref{e:Q1-res}) and
(\ref{e:Q4-res}), we obtain
\bean
	\tilde A_Q(r;s,\phi) &=& \frac r2\bigl[
		  \e{i2\phi}(-Q_1+Q_2)
		+ \e{-i2\phi}(-Q_3+Q_4) 
	\\&&\qquad{}
		- \pi\e{-i2\phi}(a_1-a_2)^2\bigr].
\eean
However, this integral should be real, so the imaginary part must be
zero:
\[
	-Q_1 + Q_2 + Q_3 - Q_4 + \pi(a_1-a_2)^2 \equiv 0.
\]
Thus, the final expression for the angle-averaged polarization
amplification function is
\bea
\tilde A_Q(r;s,\phi) &=& -r\cos2\phi(Q_1-Q_2) 
	\label{e:AQ-res}\\
	&=& -r\left(2\frac{l^2}{s^2}-1\right)(Q_1-Q_2).
	\nonumber
\eea
This also has been confirmed by numerical integration.

\eqnref{AI-res} and \eqnref{AQ-res} are the principal results of this paper.

\begin{figure}
\begin{center}
\epsfxsize=\hsize
\epsfbox{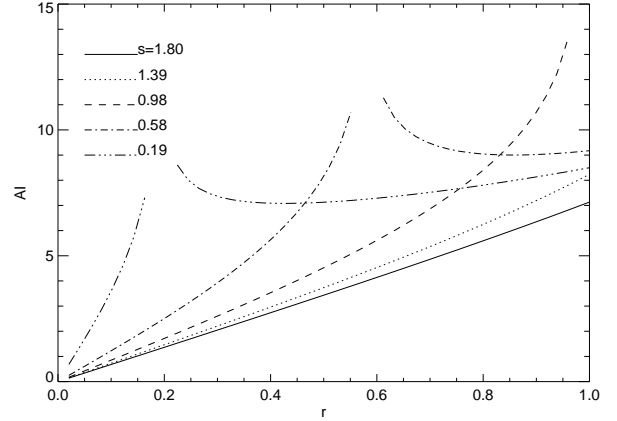}
\end{center}
\caption{\label{f:AIplot}The amplification kernel~$\tilde A_I(r;s)$,
plotted as a function of~$r$ for a selection of values of the
distance~$s$, and for an impact parameter~$l=0.1$
(see \figref{geom}).  There is a singularity along the line~$r=s$,
where the integration includes the point~$\zeta=0$.}
\end{figure}

\begin{figure}
\begin{center}
\epsfxsize=\hsize
\epsfbox{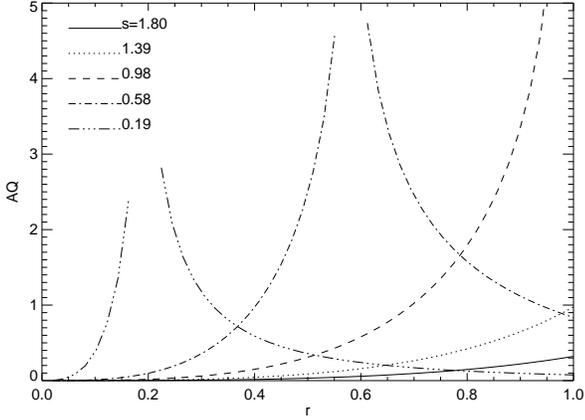}
\end{center}
\caption{\label{f:AQplot}The amplification kernel~$\tilde
A_Q(r;s,\phi)$, plotted as in \figref{AIplot}; the parameter~$\phi(s)$
is that corresponding to a straight-line transit with impact
parameter~$l=0.1$.}
\end{figure}

In \figref{AIplot},
I show the amplification function~$\tilde A_I(r;s)$, with a
singularity along the line~$r=s$.  Note particularly the breadth of
the kernel after the singularity: although the angle-averaged
amplification is close to~$1$ away from a well-defined peak at~$r=s$,
the factor of~$r$ in \eqnref{Idef} means that
there are contributions to~$I(t)$ in \eqnref{IIP} from a broad range
of~$I(r)$, a situation especially severe for cases where the source
function~$I(r)$ extends significantly beyond~$r=1$ (\textit{ie}, those
cases where~$R_*>R_E$).  The breadth and asymmetry of the kernel is
what makes the recovery of the source function so problematic.
Similarly, \figref{AQplot}
shows the amplification kernel~$\tilde A_Q(r;s,\phi)$.  This has the
same factor of~$r$ as~$A_I(r;s)$, but because the underlying
angle-averaged function is close to zero away from~$r=s$, this has a
less damaging effect, so that the polarization signal should be easier
to recover (which is fortunate, since that signal is so much harder to
detect than the intensity signal).  The kernel is still, however, both
broad and asymmetric.

\section{Amplification functions as elliptic integrals}
\label{s:elliptic}

The integrals defined by equations~(\ref{e:I1-def})
and~(\ref{e:I2-def}), and equations~(\ref{e:Q1-def})
to~(\ref{e:Q4-def}), are elliptic integrals.  Carlson~\shortcite{carlson88}
provides a set of recurrence relations to reduce such integrals to a
small set of elementary integrals, which are in turn expressed in
terms of a set of functions~$\RJ$, $\RF$ and~$\RD$ which are more symmetrical
alternatives to the traditional Legendre elliptic integrals, and which
can be evaluated efficiently, and to high accuracy, across their
entire domain.  A suitable
algorithm~\cite{gray00c} can do this (algebraically exhausting)
reduction mechanically, to obtain
\bea
\frac{\tilde A_I(r;s)}{r}-2\pi &=& I_1-I_2 \nonumber \\
&=&
2\frac{x_1-x_2}{\sqrt{x_1x_2}} \RF 
	\nonumber\\&&\qquad{}
	+ \frac43 \lambda_2(1+\lambda_1)\sqrt{x_1x_2}
	\nonumber\\&&\qquad\qquad{}\times
	\left( x_2 \RJ_1 - \frac{\RJ_2}{x_2}\right)
	\label{e:AIrhs1}\\
	&=&
	2 \frac{x_1-x_2}{\sqrt{x_1x_2}} K(\kappa)
	\nonumber\\&&\qquad{}
	+ 4\frac{1+\lambda_1}{x_2}\sqrt{x_1x_2}
	(\Pi_2-\Pi_1),
	\label{e:AIrhs2}\\
\frac{\tilde A_Q(r;s,\phi)}{-r\cos2\phi} &=& Q_1-Q_2 \nonumber\\
	&=&
	\frac{q_D\RD + q_F\RF + q_J\RJ_2}{2(x_1x_2)^{3/2}} 
	\label{e:AQrhs1}\\
	&=&
	\frac{q_eE(\kappa) + q_kK(\kappa) + q_\pi \Pi_2}{2(x_1x_2)^{3/2}}.
\label{e:AQrhs2}
\eea
Here
\bean
\lambda_1 &=& \frac{x_2-x_1}{x_1-\xb_2} > 0, \\
\lambda_2 &=& \frac{\xb_1-\xb_2}{x_2-\xb_1} > 0, \\
\kappa^2  &=& \lambda_1\lambda_2 \qquad (0 < \kappa^2 \le 1);
\eean
the coefficients~$q_a$ are
\bean
q_D &=&  \frac{(x_1+x_2)(x_1x_2+1)(x_2-x_1)^2}{3(x_1x_2-1)}, \\
q_F &=& -\frac{(x_1x_2-1)(x_2-x_1)^2}{x_2}, \\
q_J &=& -\frac{(x_2^2-1)(x_2-x_1)^3}{3x_2^2}, \\
q_e &=& -\frac{3q_D}{\kappa^2} = -(x_1+x_2)(x_1^2x_2^2-1), \\
q_k &=&  \frac{3q_d}{\kappa^2} + q_F + \frac{3q_J}{\lambda_1}
	\nonumber\\
    &=& (-x_2^3+3x_1x_2^2+x_1+x_2)(x_1x_2-1), \\
q_\pi &=& -\frac{3q_J}{\lambda_1} 
	= \frac{(x_2^2-1)(x_1x_2-1)(x_2-x_1)^2}{x_2};
\eean
we have used the abbreviations
\bean
\RF  &=& \RF (0, 1-\kappa^2,1), \\
\RD  &=& \RD (0, 1-\kappa^2,1), \\
\RJ_i&=& \RJ (0,1-\kappa^2,1,1+\lambda_i), \\
\Pi_i&=& \Pi \left(\frac\pi2, -\lambda_i, \kappa\right);
\eean
and $\Pi(\phi=\pi/2,n,k)$, $K(k)$ and $E(k)$ are the Legendre forms
of the complete elliptic integrals, as defined in \cite[8.111]{g&r},
or \cite[17.2]{a&s} (with $k=\sin\alpha$).  These are related to the
Carlson forms of the integrals by
\bean
\RJ(0,1-\kappa^2,1;1+\lambda) &=& 
        \frac{3}{\lambda}\left[
                K(\kappa) - \Pi(\pi/2, -\lambda,\kappa)
                \right]\\
\RD(0,1-\kappa^2,1) &=&
        \frac{3}{\kappa^2}[K(\kappa) - E(\kappa)] \\
\RF(0,1-\kappa^2,1) &=&
        K(\kappa).
\eean

\subsection{Singularies and asymptotic behaviour of $\tilde A_I$ and
$\tilde A_Q$}

We can write \eqnref{AIrhs2} in terms of Heuman's Lambda function
\cite[17.4.39]{a&s} as follows:
\bea
I_1-I_2 &=& 
	2\frac{x_2-x_1}{\sqrt{x_1x_2}}K(\kappa)
	\nonumber\\&&\qquad{}
	+ 2\pi\Bigl[\Lambda_0
		\bigl(\arcsin(1+\lambda_1)^{-1/2}
			\bigm\backslash\alpha\bigr)
	\nonumber\\&&\qquad\qquad{}
		- (\lambda_1\to\lambda_2)\Bigr].
	\label{e:AIrhs3}
\eea
The advantage of this is that we can now easily isolate the
singularity at~$r=s$, where we have~$x_1=\xb_1=1$, $x_2>1$, and
thus~$\kappa=1$.  The~$\Lambda_0$ function has no singularities, and the
coefficient of~$K(\kappa)$ is finite there, so the only singularity is
at~$K(\kappa=1)$ where \cite[17.3.26]{a&s}
\[
	\lim_{\kappa\to1}\left(K-\frac12 \ln\frac{16}{1-\kappa^2}\right)=0.
\]
Thus the leading order term in~\eqnref{AIrhs2} at~$r=s$ is
\bea
	\frac{\tilde A_I(r\sim s;s)}{r}-2\pi
		&=& I_1-I_2 (r\sim s) 
	\nonumber\\
		&\sim& \frac{x_2-x_1}{\sqrt{x_1x_2}}\ln\frac{16}{1-\kappa^2}.
	\label{AIlead}
\eea
We can also confirm the behaviour as~$r\to0$ and~$r\to\infty$.
For~$r>s$, we have
\be{x12rinfty}
	\frac{x_2}{x_1} = 1 + \frac12\frac{(4+s^2)(4-s^2)}{r^2}
		+ O(r^{-4})
\ee
and
\be{sqrtx12infty}
	\sqrt{\frac{x_2}{x_1}} - \sqrt{\frac{x_1}{x_2}}
		= \frac12 \frac{(4+s^2)(4-s^2)}{r^2} + O(r^{-4}).
\ee
It follows from \eqnref{x12rinfty} that both~$(1+\lambda_1)$
and~$(1+\lambda_2)$ go to~$1$ as~$r\to\infty$, so that the difference
of Lambda functions in \eqnref{AIrhs3} goes to zero.  The
factor~$\kappa\to0$ in this limit, and~$K(0)$ is finite, but the
coefficient of~$K$ goes to zero, from \eqnref{sqrtx12infty},
so~$I_1-I_2\to0$, and~$\tilde A_I\to2\pi r$, as expected.

For~$r<s$, both~$x_1$ and~$x_2$ diverge as~$r\to0$, but
$x_2/x_1=1+4/s^2+O(r^2)$, so that~$\kappa\to0$.  Both~$K$
and~$\Lambda_0$ are finite here, so that~$I_1-I_2$ does not diverge,
and the singularity is confined to the coordinates~$x_i$.

We may now move on to the difference~$Q_1-Q_2$.  After
rewriting~$\Pi_2$ in terms of~$K$ and~$\Lambda_0$, the coefficient
of~$K(\kappa)$ in \eqnref{AQrhs2} is
$(q_k+q_\pi/(1+\lambda_2))/2(x_1x_2)^{3/2}$.  This is a rather messy
expression in general, but at~$r=s$, where~$x_1=1$, its value is
$2(\sqrt{x_2}-1/\sqrt{x_2})$, so that the leading order term in
\eqnref{AQrhs2} is
\bea
	\frac{\tilde A_Q(r\sim s;s,\phi)}{-r\cos2\phi} 
		&=& Q_1-Q_2 (r\sim s)
	\nonumber\\
		&\sim& \left(\sqrt{x_2}-\frac{1}{\sqrt{x_2}}\right)
			\ln\frac{16}{1-\kappa^2}.
	\label{AQlead}
\eea
One can draw the same conclusion directly from \eqnref{AQrhs1} by
using the useful asymptotic expansions in~\cite{carlson94},
specifically relations~(26), (34) and~(44) in that paper.

Since the singularity in~$\tilde A_I$ and~$\tilde A_Q$ is no worse
than logarithmic, we may numerically evaluate integrals involving these
by using Gaussian quadrature with a log weight.

\section{Conclusion}

I have obtained analytic angular integrals of the microlensing
amplification function, for the case of a rotationally symmetric
source.  This avoids the need to use approximate 
methods to obtain this expression, and means that they can be
evaluated more efficiently than using general numerical
integrations.  Also, we are able to make analytic statements about the
leading-order behaviour of the integrals along their~$r=s$
singularity, and so use such asymptotic approximations in further
treatments.

This also means that the dependence of the observed flux on the limb-darkening
function, and of the observed polarization on the limb-polarization
function, can be expressed as integral equations.  Thus the problem of
recovering the latter from the former can be viewed as a classic
inverse problem, which can be analysed in detail using the
sophisticated techniques developed for such problems.  This is
the subject of a forthcoming paper~\cite{gray00a}.

{
\def\JApJ{ApJ}
\def\JMNRAS{MNRAS}
\def\JAA{A\&A}

}

\end{document}